\pdfoutput=1
\documentclass[aip,pof,12pt]{revtex4-1}
\usepackage{graphicx}
\usepackage{amssymb,amsmath}
\usepackage{natbib}
\usepackage{subfigure}
\usepackage{color}

\newcommand{\nom}{Nu_\omega}

\begin{document}

\title{Boundary layer dynamics at the transition between the classical and the ultimate regime of Taylor-Couette flow}
\author{Rodolfo Ostilla M\'onico$^1$, Erwin P. van der Poel$^1$, Roberto Verzicco$^{1,2}$, Siegfried Grossmann$^{3}$ and Detlef Lohse$^{1}$}
\affiliation{
$^{1}$ Physics of Fluids Group, Department of Science and Technology, Mesa+ Institute,  and J.\ M.\ Burgers Centre for
Fluid Dynamics,\\ University of Twente, 7500 AE Enschede, The Netherlands \\
$^2$Dipartimento di Ingegneria Industriale, University of Rome ``Tor Vergata'', Via del Politecnico 1, Roma 00133, Italy \\
$^3$ Fachbereich Physik, Philipps-Universit\"{a}t Marburg, Am Renthof 6, D-35032 Marburg, Germany}
\date{\today}

\begin{abstract}
Direct numerical simulations of turbulent Taylor-Couette flow are performed up to inner cylinder Reynolds numbers of $Re_i=10^5$ for a 
radius ratio of $\eta=r_i/r_o=0.714$ between the inner and outer cylinder. With increasing $Re_i$, the flow 
undergoes transitions between three different regimes:
(i) a flow dominated by large coherent structures, (ii) an intermediate transitional regime, and (iii) a flow with developed turbulence. 
In the first regime the large--scale rolls completely drive the meridional flow while in the second one the coherent structures 
recover only on average. The presence of a mean flow allows for the coexistence of laminar and turbulent boundary layer dynamics. 
In the third regime the mean flow effects fade away and the flow becomes dominated by plumes. The effect of the local driving on the 
azimuthal and angular velocity profiles is quantified, in particular we show when and where those profiles develop.
\end{abstract}

\maketitle

Taylor-Couette flow (TC), the flow between two independently rotating coaxial cylinders, and Rayleigh-B\'enard flow (RB), the 
thermally driven flow in a fluid cell heated from below and cooled from above, are twin flows \cite{bus12}, with analogous 
global balances \cite{eck07}. The temperature difference between the plates in RB flow corresponds to the different rotation rates of the 
inner and outer cylinders in TC flow (hereafter referred to as drivings of the flow).

Kraichnan \cite{kra62} postulated that RB flow would reach an asymptotic or ``ultimate'' state if the driving is sufficiently large.
In this ultimate regime, turbulence is fully developed throughout both bulk and boundary layers. In that regime, heat transport, which for less driving is
limited by the laminar boundary layers, has no explicit, unit-wise dependence of viscosity.
The ultimate-regime scaling laws relating heat transfer and flow driving are expected to be extendable to arbitrarily large 
Rayleigh numbers, like those present in both geo- and astrophysical flows.

Indeed, such transitions in the scaling laws expected in this ultimate state have recently 
been found in experiments for heat transport and wind in RB flow \cite{roc10,he12} and angular velocity transport and wind
in TC flow \cite{lew99,gil12,hui12}, confirming the existence of the ultimate regime, 50 years after Kraichnan's prediction.

RB and TC flow are expected to show the transition to the ultimate regime when the boundary layers are sheared strongly enough so that they 
undergo a shear-instability and become turbulent \cite{gro11,gro12,gil12}.  
A universal characteristic of turbulent boundary layers is the so-called law-of-the-wall \cite{pra25,kar30,tow76}. For wall distances
much larger than the internal length scale and much smaller than the outer length scale,
the mean velocity profile has a logarithmic dependence on the distance to the wall. 
This region has been found in many different types of flows, both experimentally \cite{wei89} and numerically \cite{kim87}. 
For further details on the empirical evidence for the universality of this region, we refer the reader to recent reviews by Smith, McKeon \&
Marusic \cite{smi11} and Jimenez \cite{jim12}. 

As in other flows, in TC and RB turbulent boundary layers are expected to produce this characteristic logarithmic signature,
not only in the mean velocity, but, in RB, also in the mean temperature profile \cite{gro12}. 
These log-layers, which extend significantly into the flow, have been experimentally detected \cite{hou11} and measured \cite{hui13} in TC.
Later, in a theoretical work, Grossmann et al. \cite{gro14} pointed out that in addition to the axial velocity, it should be the (properly
shifted) \emph{angular} velocity which should be closest to a log-behavior, rather than the \emph{azimuthal} velocity, but both should show
curvature corrections.
In RB flow, logarithmic mean temperature profiles were measured \cite{ahl12} and theoretically
accounted for \cite{gro12} beyond the onset of the ultimate regime, 
which suggested that indeed, in the ultimate regime, the boundary layers are fully turbulent \cite{ahl12}. 
However, unexpectedly, log--layers (in the bulk) were also found \emph{below} the ultimate transition in the global scaling laws \cite{ahl12}. 

In view of all these findings we readdress the mechanism of log--layer formation since at those low drivings of the flow
prior to the BL transition the shear alone is not large enough to generate turbulent boundary layers.
Based on direct numerical simulations and on \emph{local} (i.e., $z$ dependent) velocity profiles
 will develop a novel viewpoint of the laminar-turbulence transitions, 
linking them not only to either bulk or boundary layer transitions as has been hitherto done, but also to the interaction between 
bulk and boundary layers.
This viewpoint will lead to the following picture: 
for low driving, the boundary layers are of Prandtl-Blasius type (PB) \cite{sch79}. 
For the strongest driving, in the ultimate regime, the boundary layers 
are turbulent and coherence is lost in the bulk. Both these regimes were predicted within the unifying theory
of Eckhardt, Grossmann and Lohse\cite{eck07}. 
In between these two regions we will identify yet another regime which we denote as ``transitional regime'', where PB-type and
turbulent boundary layers coexist. In this transitional regime, the local scaling laws do not yet show the characteristic increase of 
transport which is seen in the ultimate regime.

Direct numerical simulations (DNS) of the Taylor-Couette system have been performed by numerically integrating the Navier--Stokes equations using a second--order accurate 
finite--difference code \cite{ver96}. Simulations give access to the complete flow field, and this allows for the analysis of 
the different flow regimes and for an identification of the mechanisms which lead to the transitions. The flow 
is driven by the rotation rate difference of the
inner and outer cylinders, which can be expressed in non-dimensional form as the Taylor number $Ta=\frac{1}{4}\sigma(r_o-r_i)^2(r_o+r_i)^2\omega_i^2\nu^{-2}$. 
Here $r_o$ and $r_i$ are the radii and $\omega_o$ and $\omega_i$ the angular velocities of the outer and inner cylinder, respectively.
$Ta$ is the analogue of the Rayleigh number in RB while 
$\sigma=[(r_o+r_i)/(2\sqrt{r_ir_o})]^4$ can be considered as a geometrical 
Prandtl number, which gives the relationship between the ``wind'' (i.e. $u_r$ and $u_z$) boundary layer, and the
angular velocity boundary layer \cite{eck07}. The response of the system is the torque required to drive the cylinders. 
It can be nondimensionalized as a pseudo--Nusselt number\cite{eck07} $\nom=T/T_{pa}$  where $T$ is the torque and $T_{pa}$ 
the torque required to drive the cylinders when the flow is purely azimuthal. We also define the non-dimensional radius $\tilde{r}$ to be
$\tilde{r}=(r-r_i)/(r_o-r_i)$ and the non-dimensional height $\tilde{z}$ to be $\tilde{z}=z/(r_o-r_i)$.

For the present simulations, the geometry of the system will have a fixed radius--ratio of $\eta=r_i/r_o=0.714$, and periodic boundary
conditions in the axial direction. For the smaller $Ta$, data will be taken from Ostilla et al. \cite{ost13}. These
originate from a simulation of the full domain, with $\Gamma=L/(r_o-r_i)=2\pi$, where $L$ is the axial domain length. 
For the largest $Ta$, a ``reduced geometry'' has been used: This is done in the spirit
of Brauckmann and Eckhardt\cite{bra13}, where it is shown that (i) one pair of vortices in 
the axial direction gives the same first order statistics as three pairs 
and (ii) forcing the system to have a rotational symmetry of order $6$ does not affect the mean flow statistics. 
Accordingly, for the largest $Ta$, i.e. $Ta>10^8$, the aspect ratio has been taken as $\Gamma=2\pi/3$
and a rotational symmetry of order $6$ is imposed on the system. This reduces the computational requirements by 
a factor of $20$ and allows us to perform the largest Ta-range simulations.

For the present study, the outer cylinder will be kept at rest, and only the inner cylinder will drive the flow. A uniform grid is used
in the azimuthal and axial directions, while a Chebychev-type clustering near the cylinders is used in the radial direction.  For temporal
convergence, two criteria must be satisfied: simulations are run until the difference between the time-averaged torque of inner and outer cylinder 
is less than $1\%$, and the average between these two values is taken for $\nom$. The simulations are then run for at least $40$ large eddy
turnover times ($(r_o-r_i)/(r_i\omega_i)$). 

\begin{figure}[t!]
 \centering
 \includegraphics[width=0.65\textwidth]{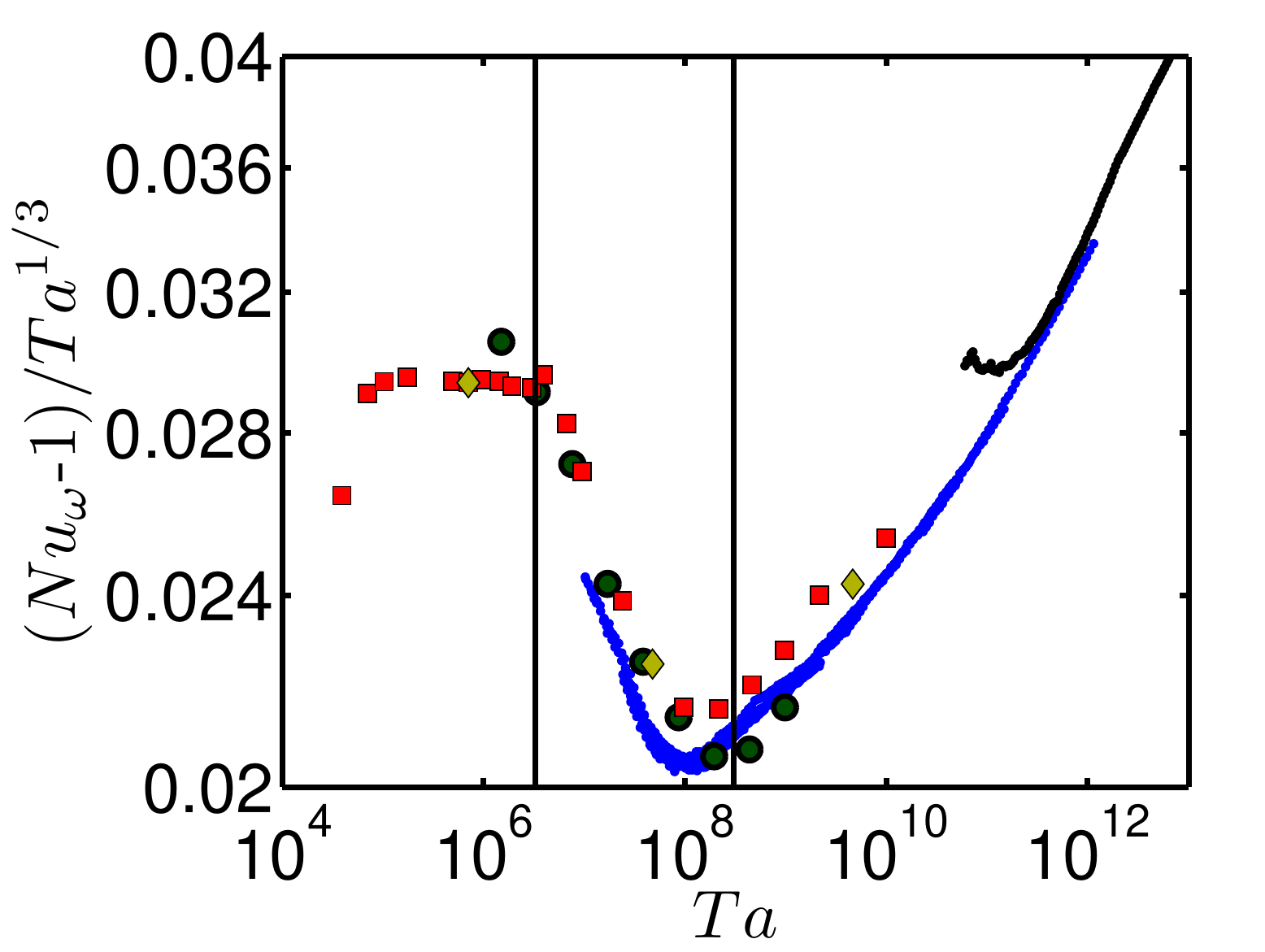}
 \caption{(color online) Compensated $\nom$ vs $Ta$ for $\eta=0.714$. The data are from experiments 
(blue dots \cite{lew99}, black dots \cite{vee14}) 
and numerical simulations (black circles \cite{bra13}, and red squares for 
older \cite{ost13} and present work). The three
dark yellow diamonds are also numerical simulations from present work and 
correspond to the cases shown in Figure \ref{fig:instq1}. The thick black
vertical lines indicate the transition between the regimes. 
}
\label{fig:comparsion}
\end{figure}

Figure \ref{fig:comparsion} shows the relationship between $Ta$ and $\nom-1$ including existing experimental
\cite{lew99,vee14} and numerical data \cite{bra13,ost13} and those of the present study (see Appendix for 
details of the numerics, including those on the numerical resolution).
After the onset of Taylor vortices at $Ta\approx10^4$ and up to $Ta\approx3\cdot10^6$, 
a laminar regime with a scaling law of $(\nom-1)\sim Ta^{1/3}$ is found. With increasing $Ta$, 
time dependence of the flow sets in, and the large--scale coherent 
structures break up into smaller structures. The vortex topology changes
from large scale rolls in the center of the flow to hairpin vortices near the boundary layer \cite{ost13}. 
The transition to the ultimate regime appears 
around $Ta\approx3\cdot10^8$, when the exponent of the scaling law $(\nom-1)\sim Ta^\alpha$
grows to values $\alpha>1/3$. As mentioned previously, this transition has been 
linked to a transition towards the turbulent state of the boundary layer \cite{gro11,gro12,gil12}.

However, the situation is more complicated. Between the laminar and ultimate regimes there is a 
transitional regime in which a mixture of both laminar and turbulent boundary layers exists. 
Analyzing the simulated flow shows that the large--scale wind generated by the coherent vortex pairs interacts with the angular
velocity boundary
layers, and induces regions where an axial pressure gradient is present. 
This pressure gradient is either favorable, and the flow is accelerated, or adverse, and the flow is decelerated. 
In the favorable pressure gradient case, the boundary layer tends to remain laminar even for intense shear 
rates owing to the stabilizing action of the pressure gradient. We wish to emphasize that this pressure
gradient comes from the \emph{wind} boundary layer, i.e. that one of the axial velocity, and acts on the $\omega$-boundary layer.
Only when $Ta$ is large enough such that the large scale vortices are weakened and eventually fade away, the boundary layers 
can be turbulent all over the axial extent, giving rise to the ultimate regime. 
This can be understood because the shear Reynolds number $Re_s$ due to the wind scales as 
$Re_s\sim\sqrt{Re_w} \sim \sqrt{Re_i} \sim Ta^{1/4}$ (cf. Ostilla
et al.\cite{ost13}) where
$Re_w$ is a ``wind''-Reynolds number, while the strength of the driving scales as $Re_i \sim Ta^{1/2}$. Therefore, in TC flow, with 
increasing $Ta$, the direct driving will eventually dominate the wind shear, and the plume growing regions will extend.

\begin{figure}[t!]
  \centering
  \includegraphics[trim=4cm   0cm 4cm 0cm,clip=true,height=6cm]{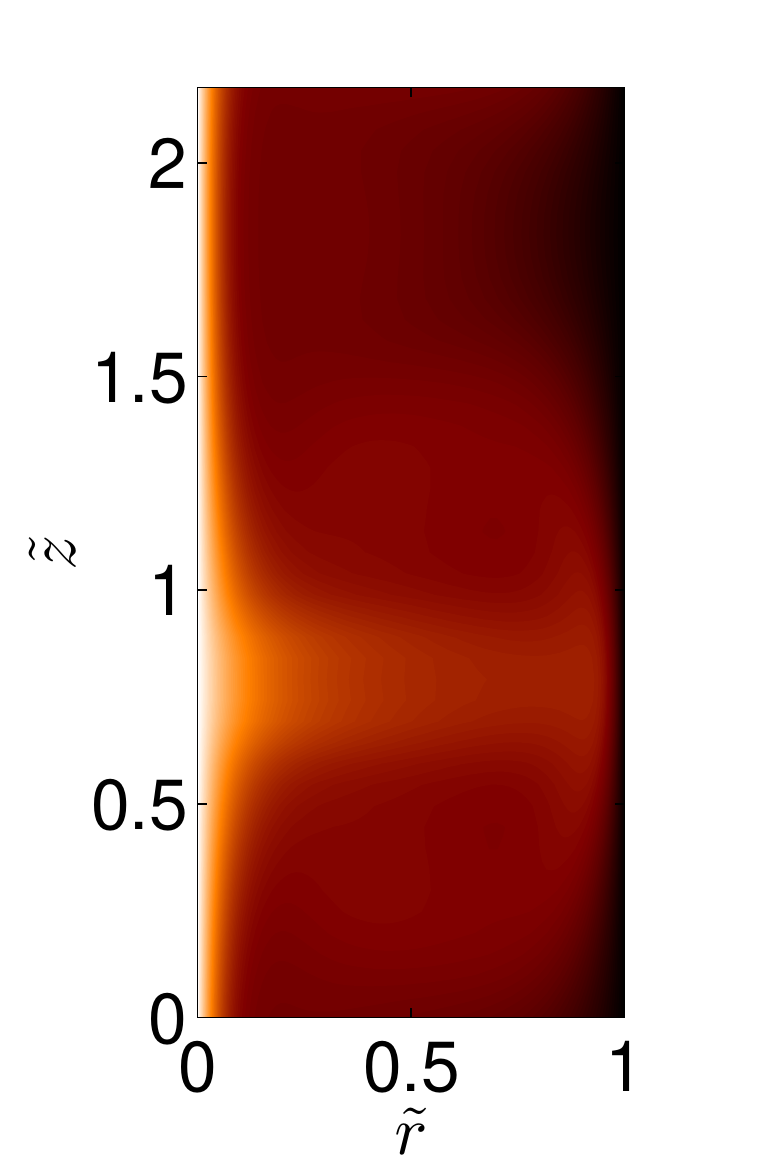}
  \includegraphics[trim=4cm   0cm 4cm 0cm,clip=true,height=6cm]{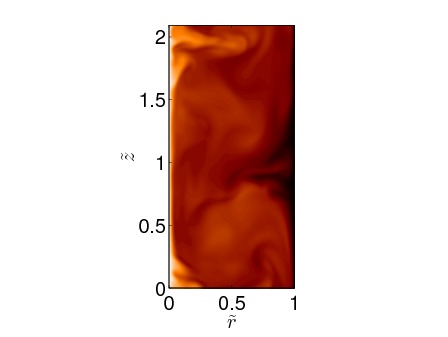}
  \includegraphics[trim=3cm   0cm 4cm 0cm,clip=true,height=6cm]{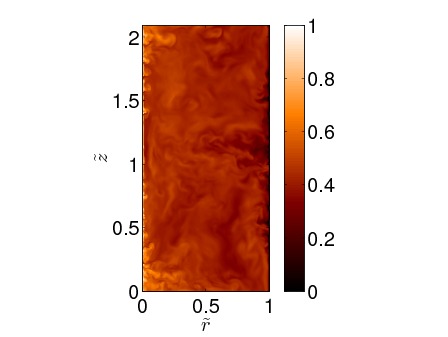}
  \caption{(color online)
Contour plots of the instantaneous azimuthal velocity field $u_\theta$ for (a) 
$Ta = 7.04\cdot10^5$ (b) $Ta = 4.77\cdot10^7$, and (c) $Ta = 4.63\cdot10^9$.
In the left snapshot, the flow is laminar. The middle snapshot is in the transition region, where laminar regions in
the boundary layer coexist with turbulent, plume ejection regions. The right-most snapshot shows fully turbulent boundary layers 
in which plumes have no preferential ejection regions. The corresponding movies are enhanced online.
}
\label{fig:instq1}
\end{figure}

Figure \ref{fig:instq1} (enhanced online) shows 
three contour plots of the instantaneous azimuthal velocity $u_\theta$ in a meridional plane (i.e., a constant $\theta$ cut)
for three values of $Ta$, in the laminar (a), transitional (b), and ultimate regime (c). The left-most panel
shows a stationary flow field in the laminar regime. The axial structure in the azimuthal velocity is produced
by the Taylor vortices. The center panel shows some plumes in the flow which are not ejected from the entire boundary layer but from 
preferential positions where the pressure gradient is adverse and separation of the boundary layer is favored.
Finally, the right--most panel shows a vanishing mean flow that, due to the negligible wall pressure gradient in the 
axial direction, allows plumes to be ejected from all over the boundary layer. 

\begin{figure}[t!]
  \centering
  \includegraphics[trim=0cm 0cm 0cm 0cm,clip=true,height=6cm,angle=-0]{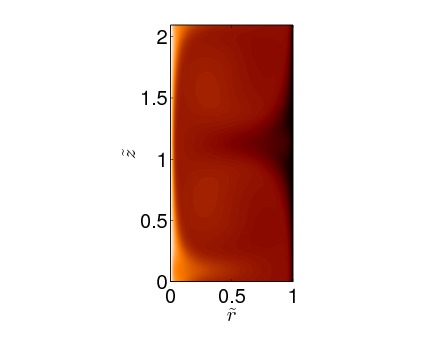}
  \includegraphics[trim=0cm 0cm 0cm 0cm,clip=true,height=6cm,angle=-0]{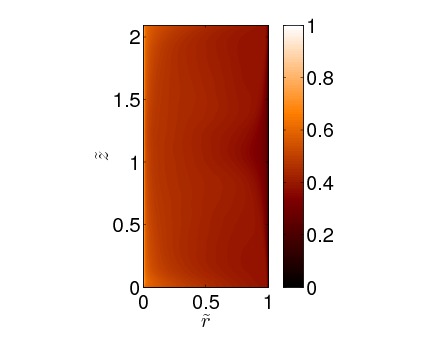}
  \caption{(color online)
Contour plots of the azimuthally- and time averaged azimuthal velocity field $\langle u_\theta \rangle_{t,\theta}$ for (a) $Ta = 4.77\cdot10^7$,
and (b) $Ta = 4.63\cdot10^9$. In the left snapshot, there are some remnants of large-scale structures, indicating preferential
ranges for plume ejection. 
In the right panel, this large scale structure is largely washed away by the wind, 
but a small structure of about $5\%$ variation still exists.}
\label{fig:meanq1}
\end{figure}

The weakening of the coherence of the wind for increasing $Ta$ can be seen in 
figure \ref{fig:meanq1}, showing two contour plots of azimuthal- and time-averaged azimuthal velocity $\langle u_\theta\rangle_{t,\theta}$ 
for two values of $Ta$. 
The left panel is for $Ta=4.77\cdot10^7$, in the transitional regime. The time averaging reveals the underlying large--scale structures;
plumes are ejected from the inner cylinder boundary layer at the preferential positions where the pressure gradient is adverse.
These plumes travel to the outer cylinder and impact the BL at preferential positions.
Similar dynamics occurs at the outer cylinder, where plumes are ejected from preferential positions, leading
to plume impacts on the inner cylinder at preferential positions.
In between the ejection and impact spots  there are ``quiet'' regions, where the wind shears the BL, but its dynamics remains laminar.
The right panel, for $Ta=4.63\cdot10^9$, corresponds to a flow field in the ultimate regime. Some structure is still present, 
but its strength is negligible  and plumes are ejected from all over the surface.

\begin{figure}[t!]
  \centering
  \includegraphics[trim=0cm 0cm 0cm 0cm,clip=true,height=6cm,angle=-0]{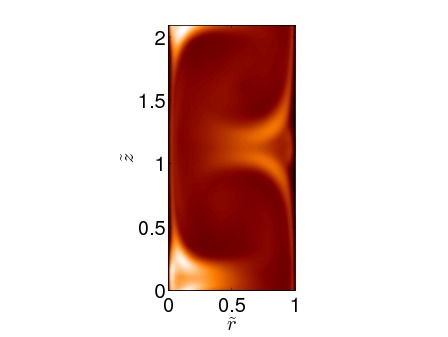}
  \includegraphics[trim=0cm 0cm 0cm 0cm,clip=true,height=6cm,angle=-0]{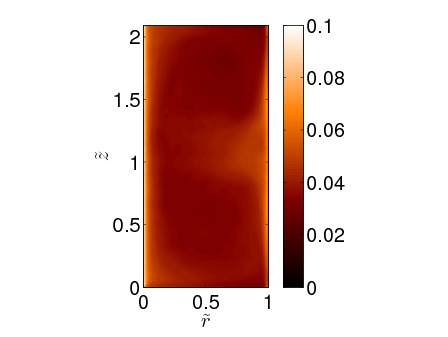}
  \caption{(color online) 
Contour plots of the azimuthal- and time-averaged r.m.s. of the velocity fluctuations $u^\prime_\theta$ for (a) $Ta = 4.77\cdot10^7$
and (b) $Ta = 4.63\cdot10^9$. The colour scales used is the same for both plots. 
In the left snapshot, the large fluctuations in the boundary layer are very
localized in the plume ejection spots. In the right snapshot, the large-scale structure in the r.m.s. 
of the velocity fluctuation seen for the left plot is almost gone,
and fluctuations (slightly smaller in absolute value) are distributed over the axial extent of the boundary layer. } 
\label{fig:rmsq1}
\end{figure}

Two contour plots of azimuthally- and time-averaged r.m.s. of the velocity fluctuations
$u^\prime_\theta = (\langle u_\theta^2 \rangle_{t,\theta} - \langle u_\theta\rangle_{t,\theta}^2)^{1/2}$ for two
values of $Ta$ can be seen in figure \ref{fig:rmsq1}.
The left panel shows the transitional regime ($Ta=4.77\cdot10^7$) where the fluctuations occur
in very localized regions which can be associated to the plume ejection spots. Outside this region, 
the fluctuations, especially in the boundary layer, are much less. 
The right panel shows the ultimate regime ($Ta=4.63\cdot10^9$).
In that regime the fluctuations are not localized and are present in the entire boundary layer.

To define the BL regions more precisely, the following criteria are used: ``wind sheared'' and ``plume-impacting/ejecting''
profiles will be identified from the axial coordinate where $\max_r |\langle u_z\rangle_{t,\theta}|$ is largest or
smallest, respectively. 
To distinguish between plume-impacting and plume-ejecting regions, the sign of $u_r$ at the mid-gap has also been measured.
When $u_r$ is positive, i.e. directed towards the outer cylinder, plumes are ejected from the inner cylinder and impact on the outer cylinder and vice versa.

\begin{figure}[ht]
    \centering
\includegraphics[trim=0cm 0cm 0cm 0cm,width=0.47\textwidth,angle=-0]{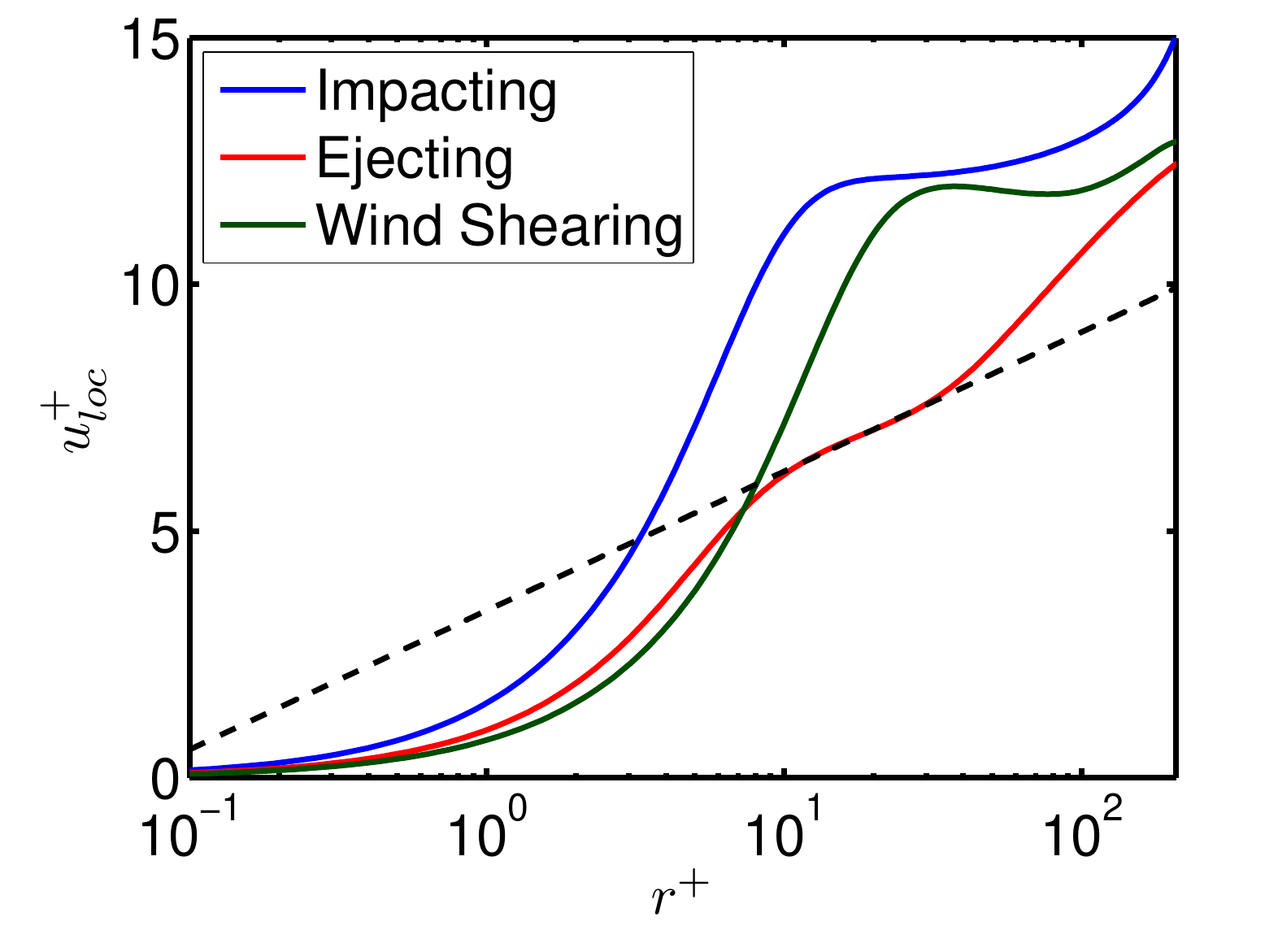}
\includegraphics[trim=0cm 0cm 0cm 0cm,width=0.47\textwidth,angle=-0]{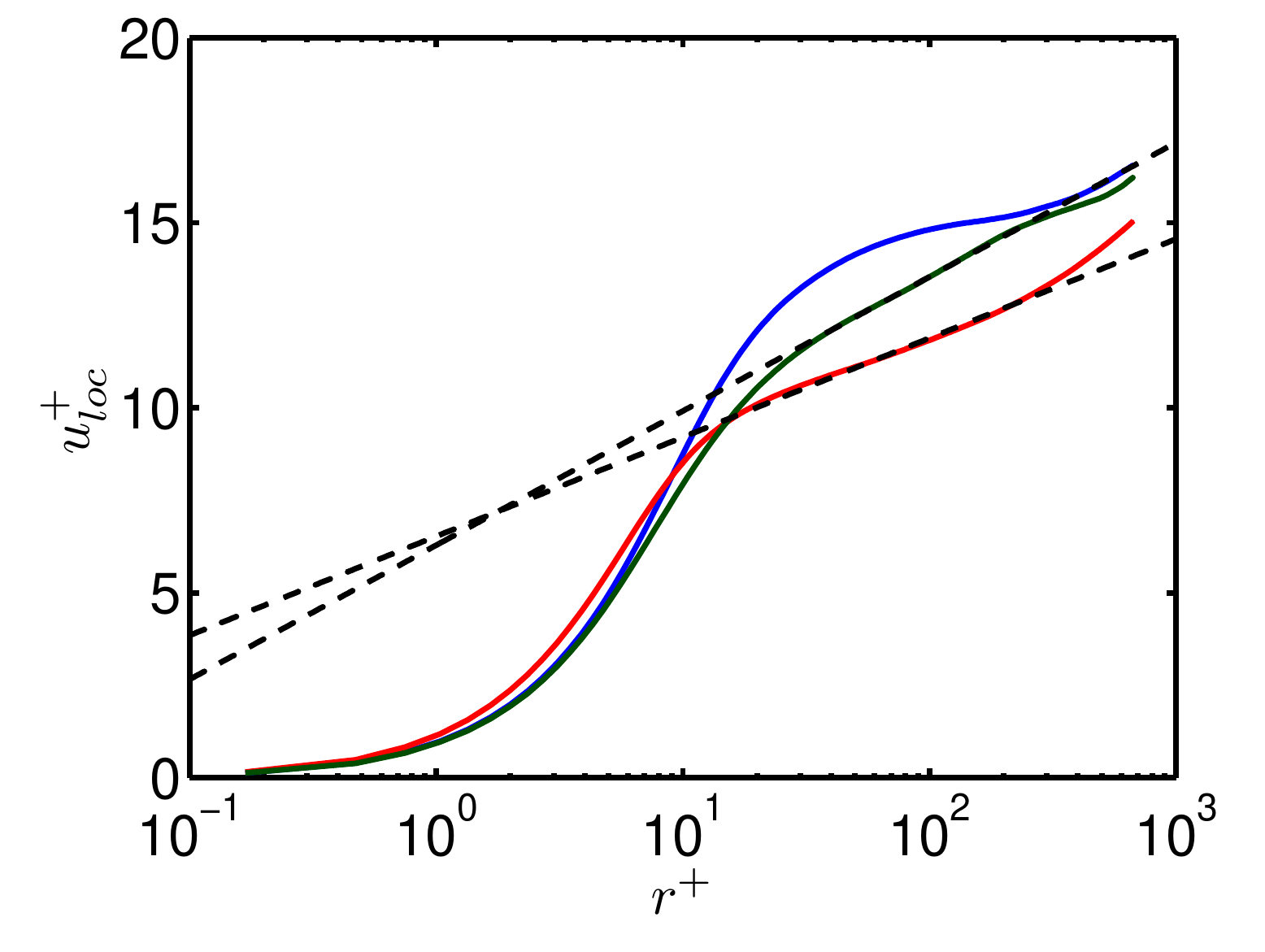}\\
\includegraphics[trim=0cm 0cm 0cm 0cm,width=0.47\textwidth,angle=-0]{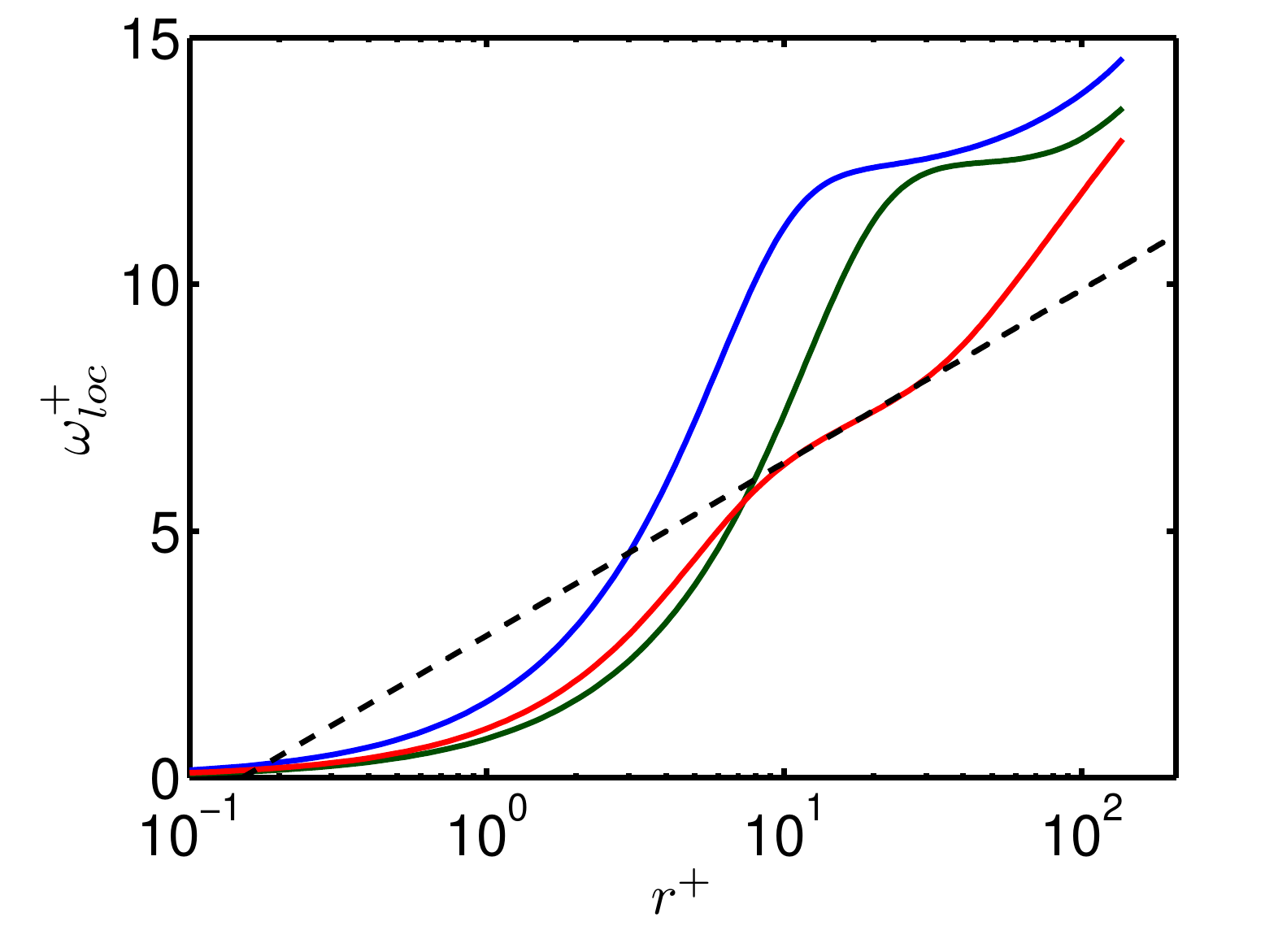}
\includegraphics[trim=0cm 0cm 0cm 0cm,width=0.47\textwidth,angle=-0]{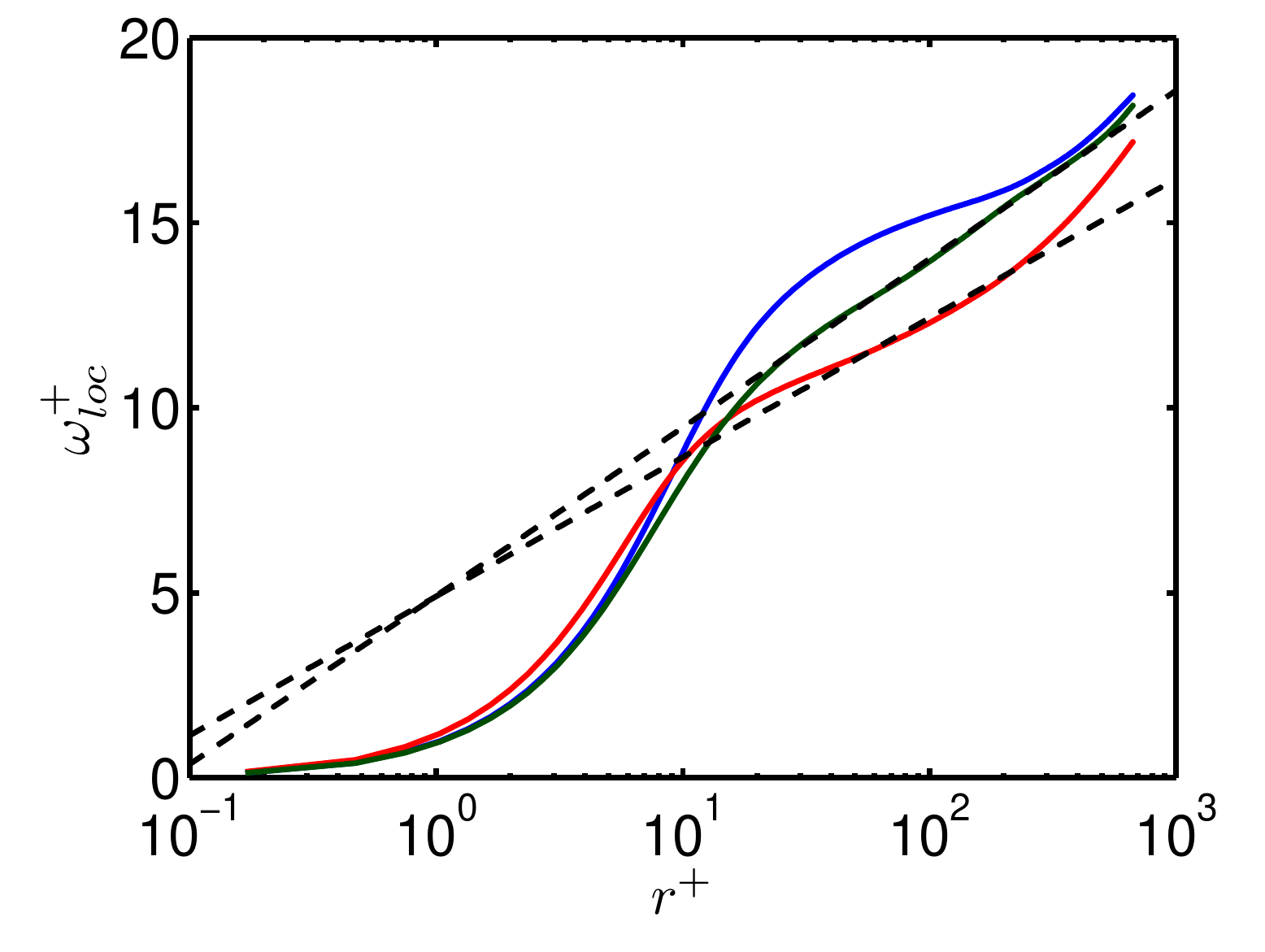}\\
    \caption{
Top panels: Local inner cylinder boundary layer profiles of $u^+_{loc}$ vs. $r^+$ for (a) $Ta = 4.77\cdot10^7$ and (b)
 $Ta = 2.15\cdot10^9$. Bottom panels: Local inner cylinder boundary layer profiles of $\omega^+_{loc}$ vs. $r^+$ 
for (c) $Ta = 4.77\cdot10^7$ and (d)  $Ta = 2.15\cdot10^9$.
In the left panels a clear distinction between the three regions can be seen. A log-layer
appears for the ejection regions at around $10<r^+<50$, but not in the other regions. This log-layer was
fitted with a straight line (in the log-lin plot) with $\kappa=0.82$ and $B=3.4$ for $u^+_{loc}$ and
$\kappa_\omega=0.65$ and $B_\omega=2.9$ for $\omega^+_{loc}$. In the right
panel, this log-layer appears also for the wind-sheared region, and now extends more into the flow, up to $r^+<300$. The difference 
in profiles between the three regions is smaller. In the ultimate regime, the ejection log-layer was
fitted with a straight line of coefficients $\kappa=0.85$ and $B=6.5$ for $u^+_{loc}$ and $\kappa_\omega=0.61$ and $B_\omega=4.90$
for $\omega^+_{loc}$, while the 
wind-sheared log-layer can be fitted with a straight line of coefficients $\kappa_\omega=0.64$ and $B_\omega=6.3$ for $u^+_{loc}$
and $\kappa_\omega=0.51$ and $B_\omega=4.9$ for $\omega^+_{loc}$.
}
\label{fig:icblseparated}
\end{figure}

Following these criteria, figure \ref{fig:icblseparated} presents
the non-dimensionalized local azimuthal velocity $u^+_{loc}\equiv(u_i-\langle u_\theta\rangle_{t,\theta})/u_*$ and the non-dimensional
local angular velocity $\omega^+_{loc}\equiv(\omega_i-\langle \omega\rangle_{t,\theta})/\omega_*$ plotted against $r^+$ for the inner cylinder boundary
layers for two values of $Ta$, one in the transitional, the other in the ultimate regime. 
For the inner cylinder, we define $u_*$, the friction velocity,
as $u_*=(\nu \langle \partial_r u_\theta(r_i)\rangle_{t,z,\theta})^{1/2}$, with $\partial_r$ the derivative normal
to the wall and $r^+$ is the non-dimensional wall distance $r^+=(r-r_i)u_*/\nu$ in wall units. Conversely, for the outer cylinder,
the frictional velocity is now defined as $u_*=-(\nu \langle \partial_r u_\theta(r_o)\rangle_{t,z,\theta})^{1/2}$, and the non-dimensional wall
distance is defined as $r^+=(r_o-r)u_*/\nu$. Finally, the frictional angular velocity $\omega_*$ is defined as $\omega_*=u_{*,i,o}/r_{i,o}$ 
for the inner and outer cylinders.

For the left panels, i.e. in the transitional regime ($Ta=4.77\cdot10^7$),
 a logarithmic region appears only for the ejection regions. In the right panels (ultimate regime, $Ta=2.15\cdot10^9$),
two log--layer ranges appear, though with different slopes, for both the wind-sheared and the plume ejection regions but not for the plume impact 
region. In the ultimate regime plumes are ejected from a larger portion of the domain, thus the distinction between these two regions becomes less sharp. 
While for the ejection regions, a logarithmic fit seems to be better for the azimuthal velocity, for the wind--sheared regions, 
logarithmic fits are better for \emph{angular} velocity.

Straight lines presenting a logarithmic Prandtl-von K\'arman--type law of the wall 
(i.e. $u^+=(1/\kappa)\log r^+ + B$ and $\omega^+=(1/\kappa_\omega)\log r^+ + B_\omega$)
were fitted through the data in the log--layer regime. This regime is expected to begin
at $r^+\approx 30$, see ref. [26]\nocite{ll87}, but begins even at a lower $r^+$ because we do not account for the axial dependence of $u_*$
and $\omega_*$ when non-dimensionalizing velocities and distances.
In the ejection regions, these fits give coefficients of $\kappa=0.82$ ($Ta = 4.77\cdot10^7$) and $\kappa=0.85$
($Ta = 2.15\cdot10^9$).  For the wind-sheared region, the $\omega$-fit gives coefficients 
of $\kappa_\omega=0.51$ and $B_\omega=4.9$ at $Ta=2.15\cdot10^9$.
In the ejection region, $\kappa$ shows at most a weak dependence on $Ta$.
Also, these coefficients deviate significantly from the classical von Karman constant $\kappa=0.41$. 
This is not surprising since the value $\kappa=0.41$ was obtained for a zero pressure gradient boundary layer. Consistently here we 
obtain values closer to $\kappa=0.41$ in the wind-sheared region where the wall pressure gradient is zero on average and switches from favorable to adverse.

What distinguishes the ultimate regime (III) from the transitional regime (II) physically? Unlike the transition from the ``laminar'' Taylor vortex
regime (I) to the ``transitional'' regime (II), which is sharp and can be associated with the onset of time dependence, the transition from the ``transitional''
regime (II) to the ultimate regime (III) resembles more a saturation process. As the plume-ejection region grows, the relative portion of the BL which is turbulent
grows, and in consequence the angular velocity transport increases. In the ultimate regime, this region cannot grow significantly any more,
and the BLs behave on average like turbulent BLs instead of Prandtl--Blasius laminar type BLs. 

\begin{figure}[ht]
 \centering
 \includegraphics[trim=0cm 0cm 0cm 0cm,width=0.48\textwidth,angle=-0]{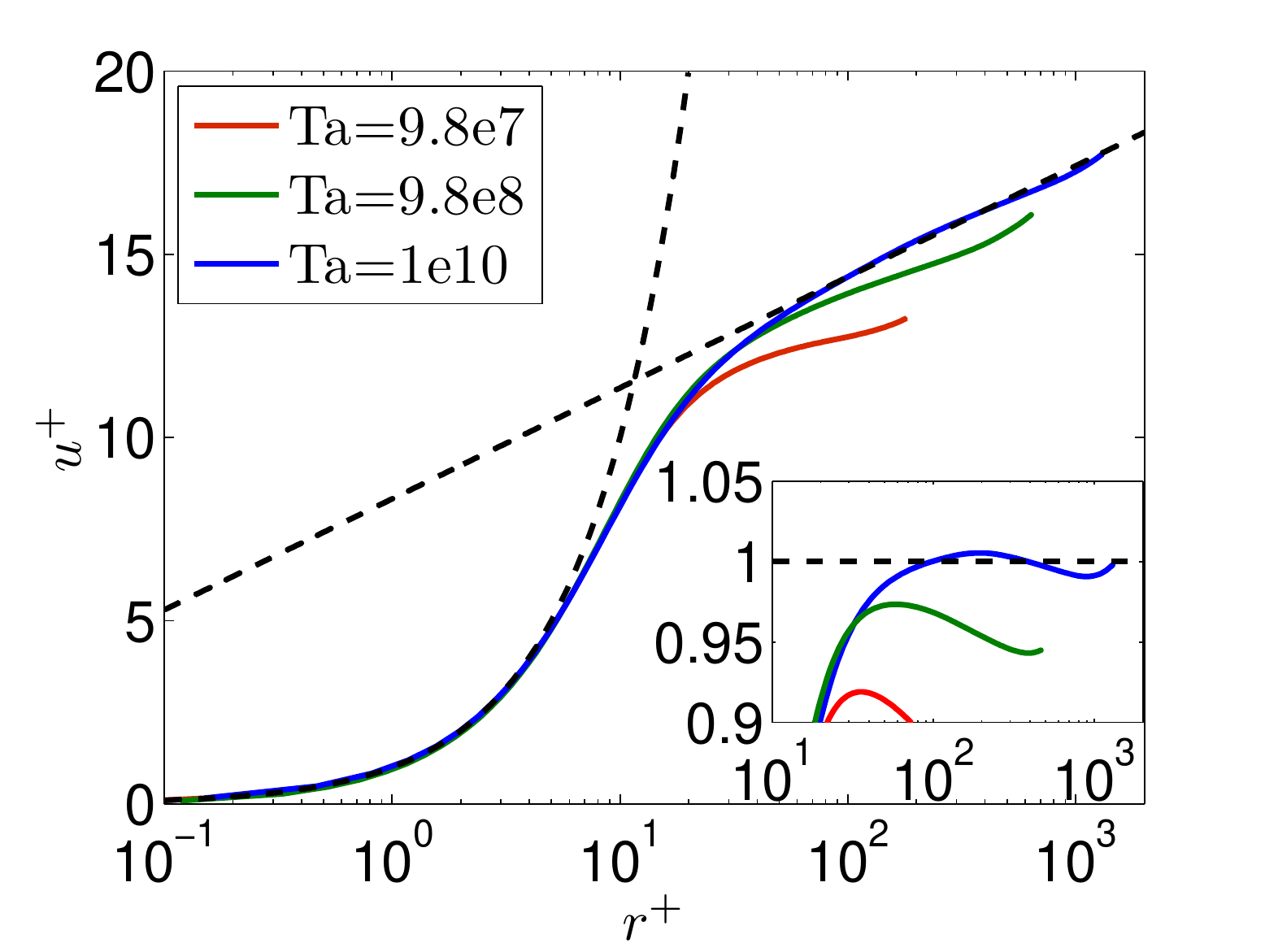}
 \includegraphics[trim=0cm 0cm 0cm 0cm,width=0.48\textwidth,angle=-0]{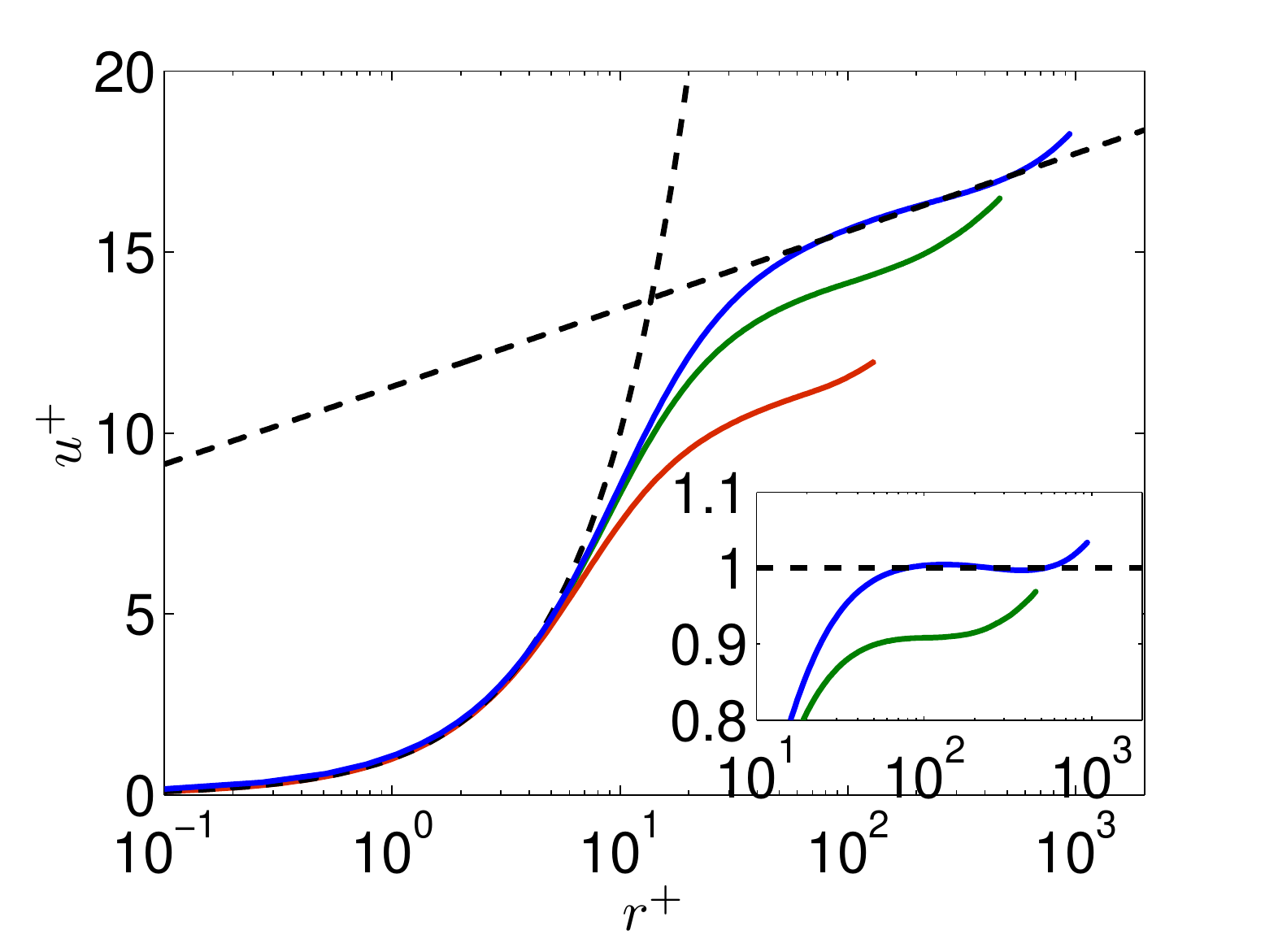}\\
\includegraphics[trim=0cm 0cm 0cm 0cm,width=0.48\textwidth,angle=-0]{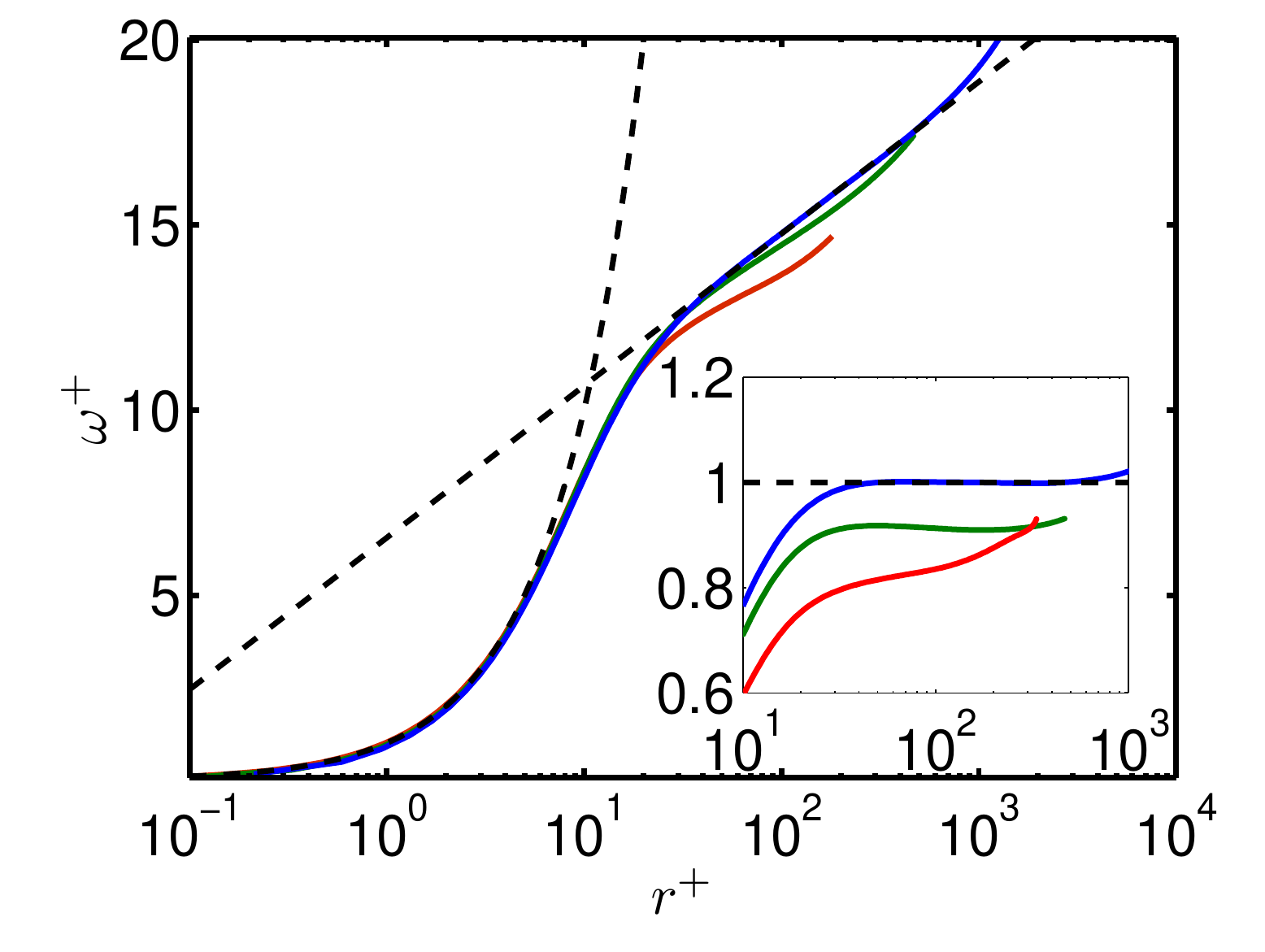}
 \includegraphics[trim=0cm 0cm 0cm 0cm,width=0.48\textwidth,angle=-0]{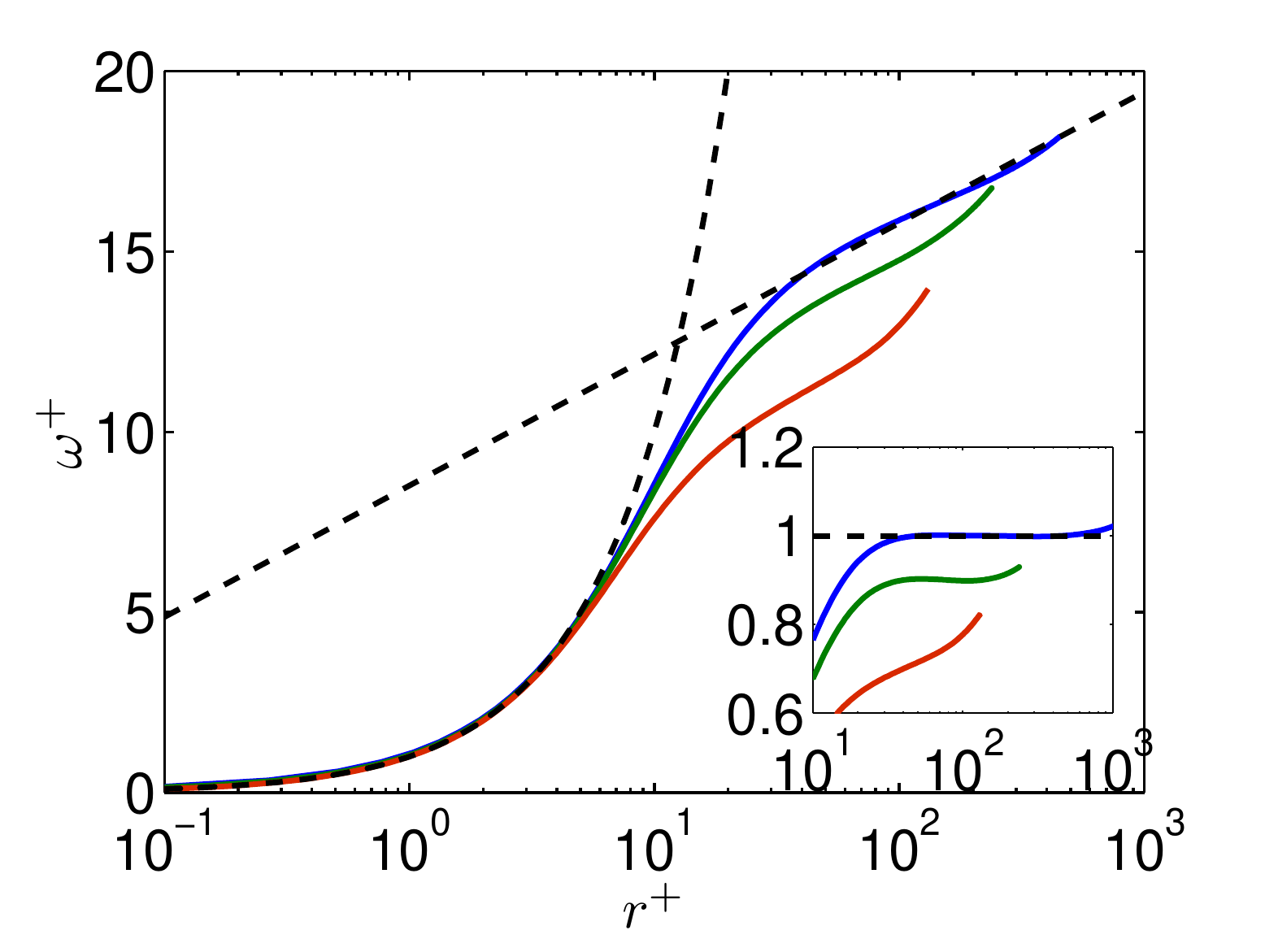}
  \caption{
Top panels: Axially averaged inner cylinder (left) and outer cylinder (right) boundary layer profiles $u^+$ versus $r^+$ for three values of $Ta$. 
Bottom panels: Axially averaged inner cylinder (left) and outer cylinder (right) boundary layer profiles $\omega^+$ versus $r^+$ for three values of $Ta$. 
Profiles are plotted from the cylinder walls to mid-gap.
No log-layers can be seen for the smallest value of $Ta$, while for the other two, in the ultimate regime,
a logarithmic signature appears, which extends for increasing values of $Ta$. 
This characterizes the transition of the whole flow to the ultimate regime. The logarithmic signatures
can be better seen when looking at the $\omega^+$ profiles rather than the $u^+$ profiles, consistent with the theory
by Grossmann et al. \cite{gro14}. Straight line 
fits for the log-region of the highest $Ta$ profile have been added, with coefficients $\kappa=0.76$ and $B=8.3$ ($u^+$), 
and $\kappa_\omega=0.56$ and $B_\omega=6.6$ ($\omega^+$) 
for the inner cylinder (left) and $\kappa=1.07$ and $B=11.3$ ($u^+$) and $\kappa_\omega=0.63$ and $B_\omega=8.5$ ($\omega^+$)
for the outer cylinder (right).
To highlight the differences across the profiles, the inset in each panel shows the three curves compensated by the logarithmic fit
of the profile with the highest $Ta$, i.e. $u^+/u^+_{fit}(Ta=10^{10})$ and $\omega^+/\omega^+_{fit}(Ta=10^{10})$.
}
\label{fig:icblmean}
\end{figure}

This can be observed in Figure \ref{fig:icblmean}, where $u^+(r^+)=(u_i-\langle u_\theta\rangle_{t,\theta,z})/u_*$ 
and $\omega^+(r^+)=(\omega_i-\langle \omega_\theta\rangle_{t,\theta,z})/\omega_*$ are plotted against
 $r^+$ for the inner and outer cylinder BLs for
three values of $Ta$. In the transitional regime, the azimuthal and angular velocity profiles lose the logarithmic behaviour when averaged axially. 
On the other hand, in the ultimate regime, the log-behaviour is maintained after averaging.

\begin{figure}[ht]
    \centering
    \includegraphics[trim=0cm 0cm 0cm 0cm,width=0.47\textwidth]{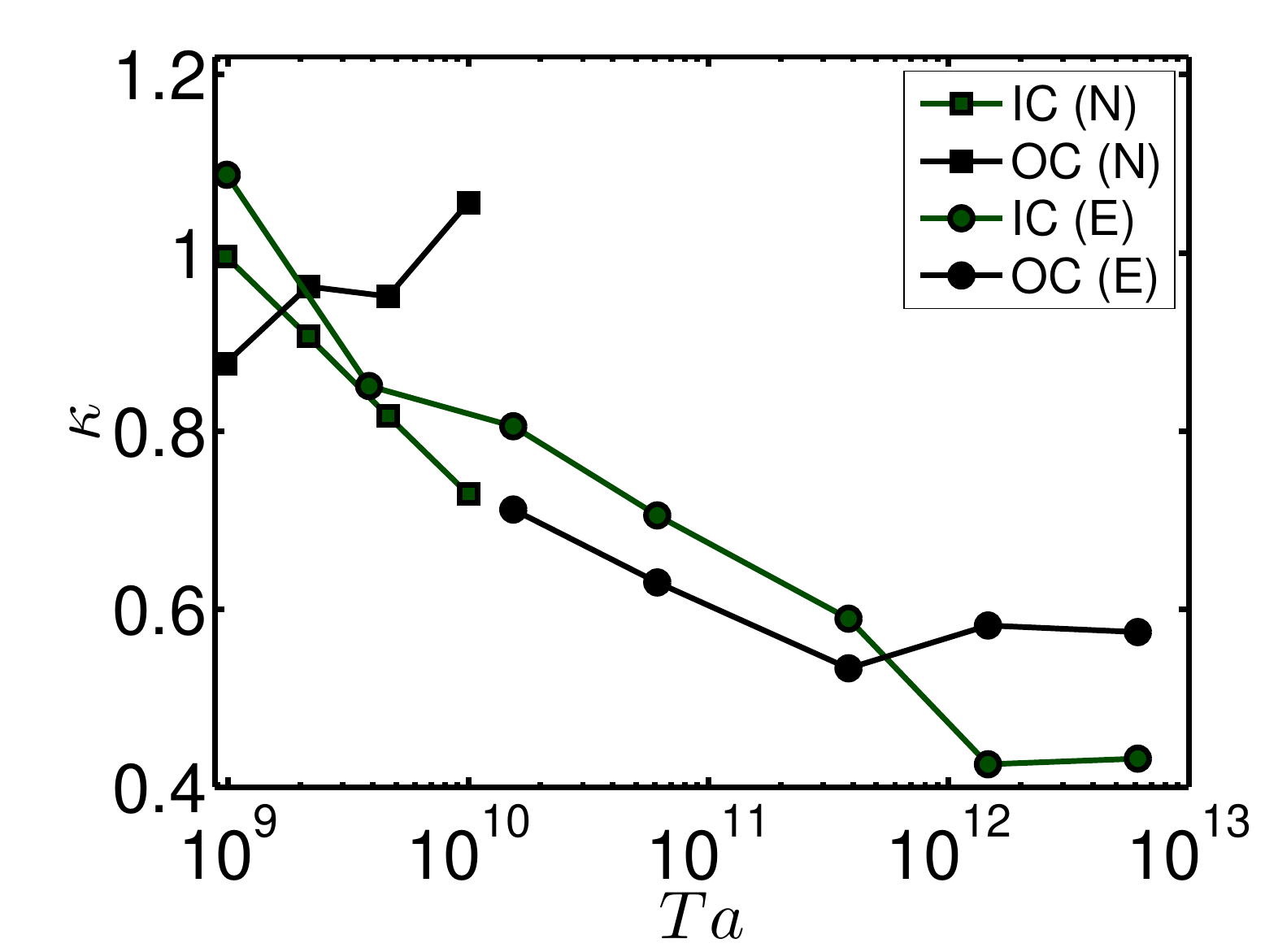}
    \includegraphics[trim=0cm 0cm 0cm 0cm,width=0.47\textwidth]{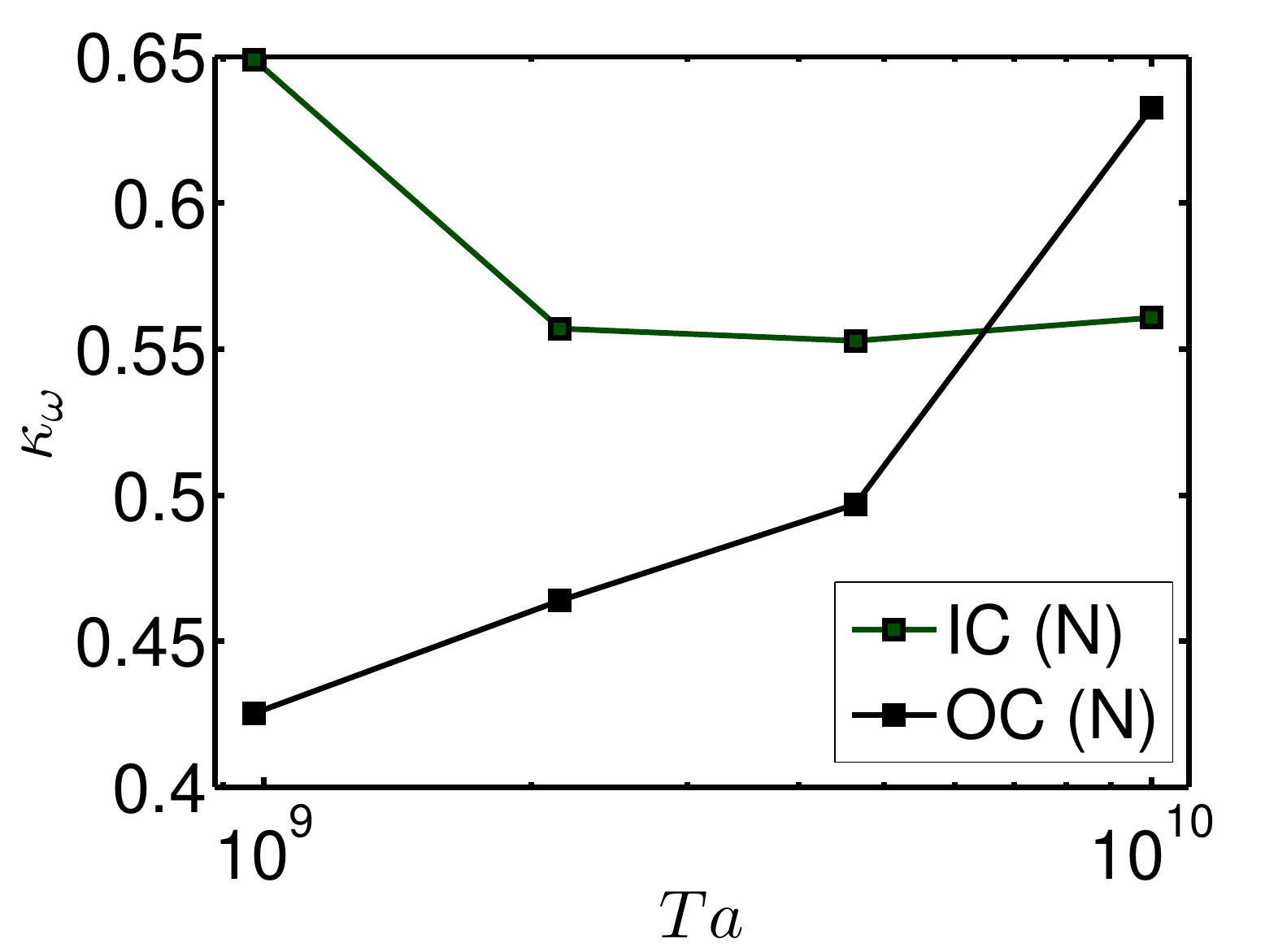}\\
    \includegraphics[trim=0cm 0cm 0cm 0cm,width=0.47\textwidth]{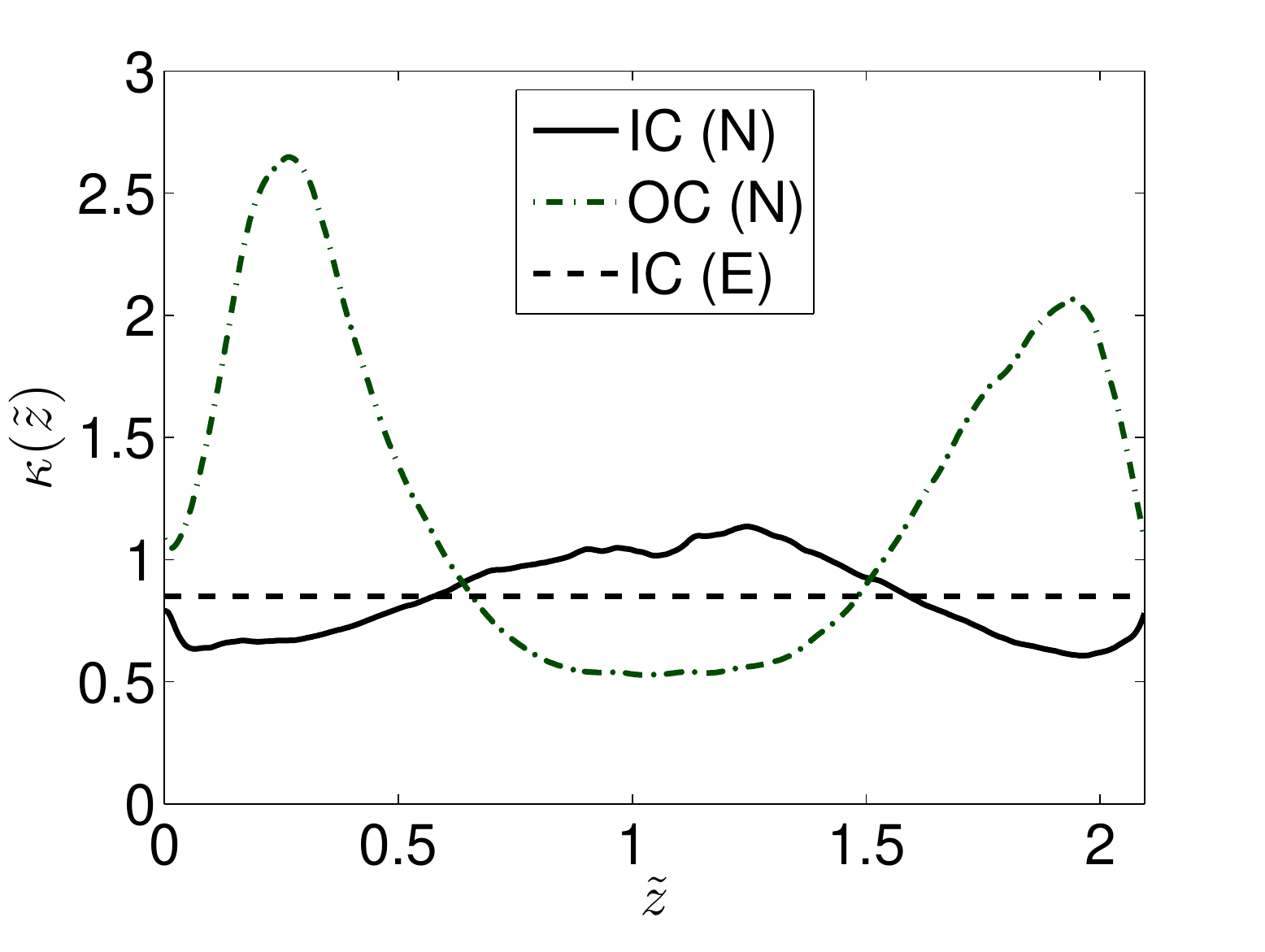}
    \includegraphics[trim=0cm 0cm 0cm 0cm,width=0.47\textwidth]{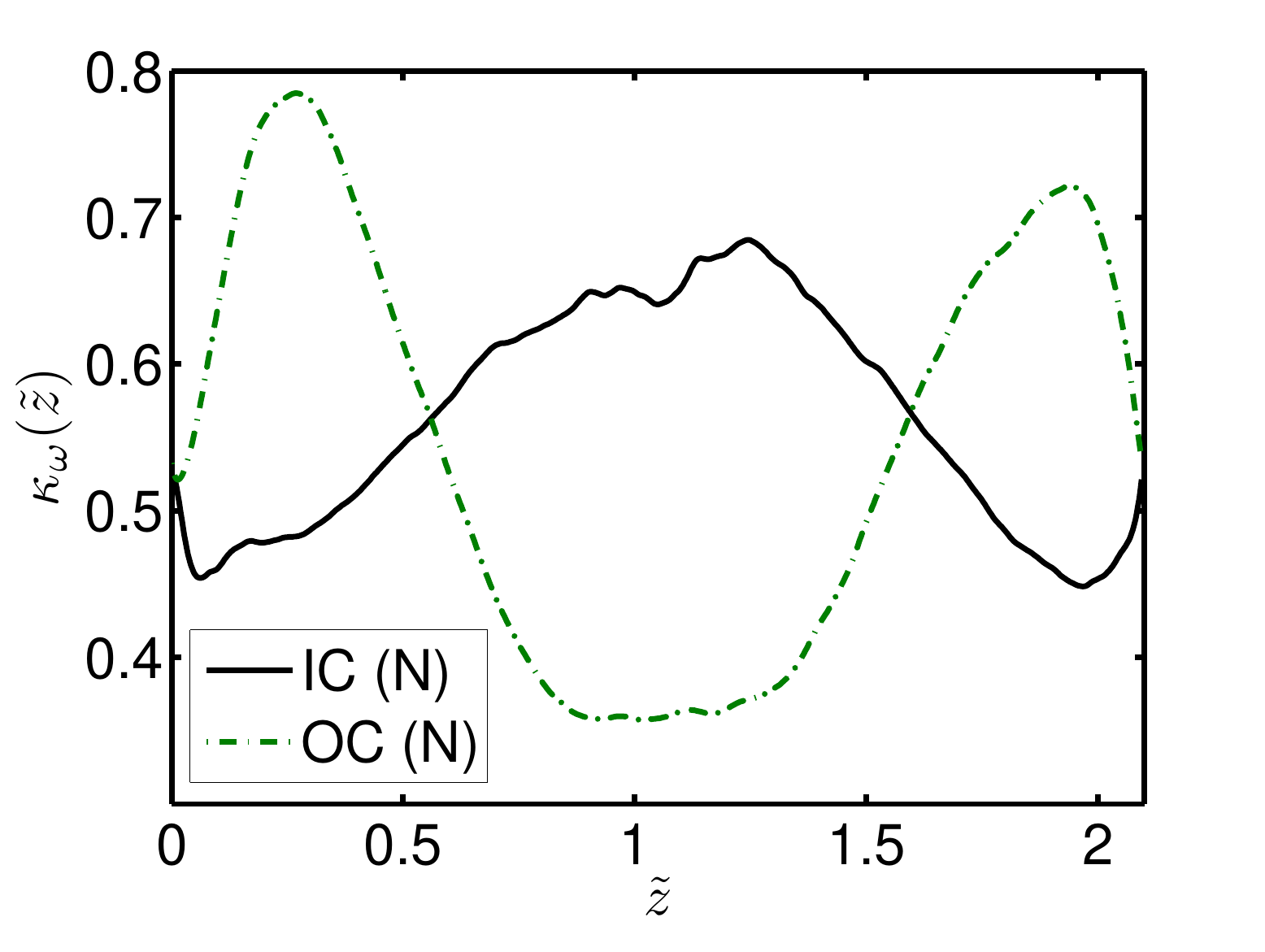}
    \caption{Top panels: dependence of the log-layer fit coefficient $\kappa$ (left) and $\kappa_\omega$ (right)
 on the driving $Ta$ for both cylinders, based on the present DNSs (N), averaged over time and in
axial direction. Experimental 
data points (E) measured for a fixed height from Huisman et al. \cite{hui13} are also shown in the 
left panel. For the inner cylinder, the inverse-slope coefficient
$\kappa$ tends to $\kappa=0.41$ with increased driving. 
For the outer cylinder, this trend is not yet visible in the numerics, probably due to insufficient driving. Bottom panels:
axial dependence of $\kappa$ (left) and $\kappa_\omega$ (right) for $Ta=4.2\cdot10^9$ for both inner and outer cylinder. 
Large fluctuations across the axial direction of the cylinder can be seen, which are smaller for the case of $\kappa_\omega$. 
On the bottom-left panel, the dashed line shows $\kappa$ at the experimental value at $Ta=3.9\cdot10^9$, closest to the numerical data,
but this value of $\kappa$ was measured at a fixed height, i.e. the cylinder's mid-height.}
\label{fig:kappafig}
\end{figure}

As also observed in experiments \cite{hui13}, a logarithmic dependence can be seen
for $u^+(r^+)$ and for $\omega^+(r^+)$ when $r^+$ is in the range $50<r^+<600$. 
Strictly speaking, not both can show a logarithmic profile, as they are related by:
$\omega^+(r^+) = u^+(r^+) + (u_\theta [r-r_i])/(ru_*)$.
As expected from the theory \cite{gro14}, the match is better
for the angular velocity profiles and not the azimuthal velocity profiles. 
The value of $\kappa$ and $\kappa_\omega$ 
may depend on the strength of the flow driving, i.e. $Ta$. 
We can quantify the dependence of $\kappa$ and $\kappa_\omega$ on the driving by fitting logarithmic
curves to $u^+(r^+)$ and $\omega^+(r^+)$, the time- and axially-averaged non-dimensional velocity profiles, 
in the range of $50<r^+<600$ (following Huisman et al. \cite{hui13}) for various $Ta$.
For the small $Ta$ cases where $r^+=600$ is farther away from the cylinder walls than the mid-gap, the mid-gap is taken as upper limit for the fit. 
The results are shown in the top panels of Figure \ref{fig:kappafig} for $Ta$ between $10^9$ and $10^{10}$. 
The residual from fitting a logarithmic profile is too large to justify a fit below $Ta\approx10^9$.
Experimental data from Huisman et al. \cite{hui13} for the $Ta$ range between $Ta=10^9$ and $Ta=6.2\cdot10^{12}$ are also plotted in the
top-left panel.

Both experimental and numerical data show the same trend for the lower drivings, i.e.  
for the lower values of $Ta$ the obtained values for $\kappa$ and $\kappa_\omega$ deviate from the classical $0.41$ von Karman constant. 
This might be due to still not large enough
driving and/or to the remnants of the large--scale structures, which is itself a consequence of this insufficient driving. Indeed, in experiments
up to $Ta=6.2\cdot 10^{12}$ is achieved, and a convergence of $\kappa(Ta)$ to a value of $0.40$ can be seen, see figure \ref{fig:kappafig}, top-left
panel. 
However, when looking at the values of $\kappa$ itself, deviations between experimental and numerical data can be seen. 
Discrepancies are probably caused by the axial dependence of $\kappa$, as the
experimental values of $\kappa$ are taken at a fixed height at the mid-cylinder, while the numerical values
originate from an axially averaged azimuthal velocity.

To further quantify this statement, the axial dependence of $\kappa$ and $\kappa_\omega$ for $Ta=4.2\cdot10^9$ is shown in the bottom panels
of figure \ref{fig:kappafig}. Large variations across the axial direction of the cylinder can be seen, which are smaller
in the case of $\kappa_\omega$.
For the bottom-left panel, the experimental value of $\kappa$ at $Ta=3.9\cdot10^9$ is plotted as a dashed line. This value is measured at a fixed
height, which can correspond to any value of the axial coordinate in the numerical domain.

\begin{figure}[ht]
    \centering
    \includegraphics[trim=0cm 0cm 0cm 0cm,width=0.48\textwidth,angle=-0]{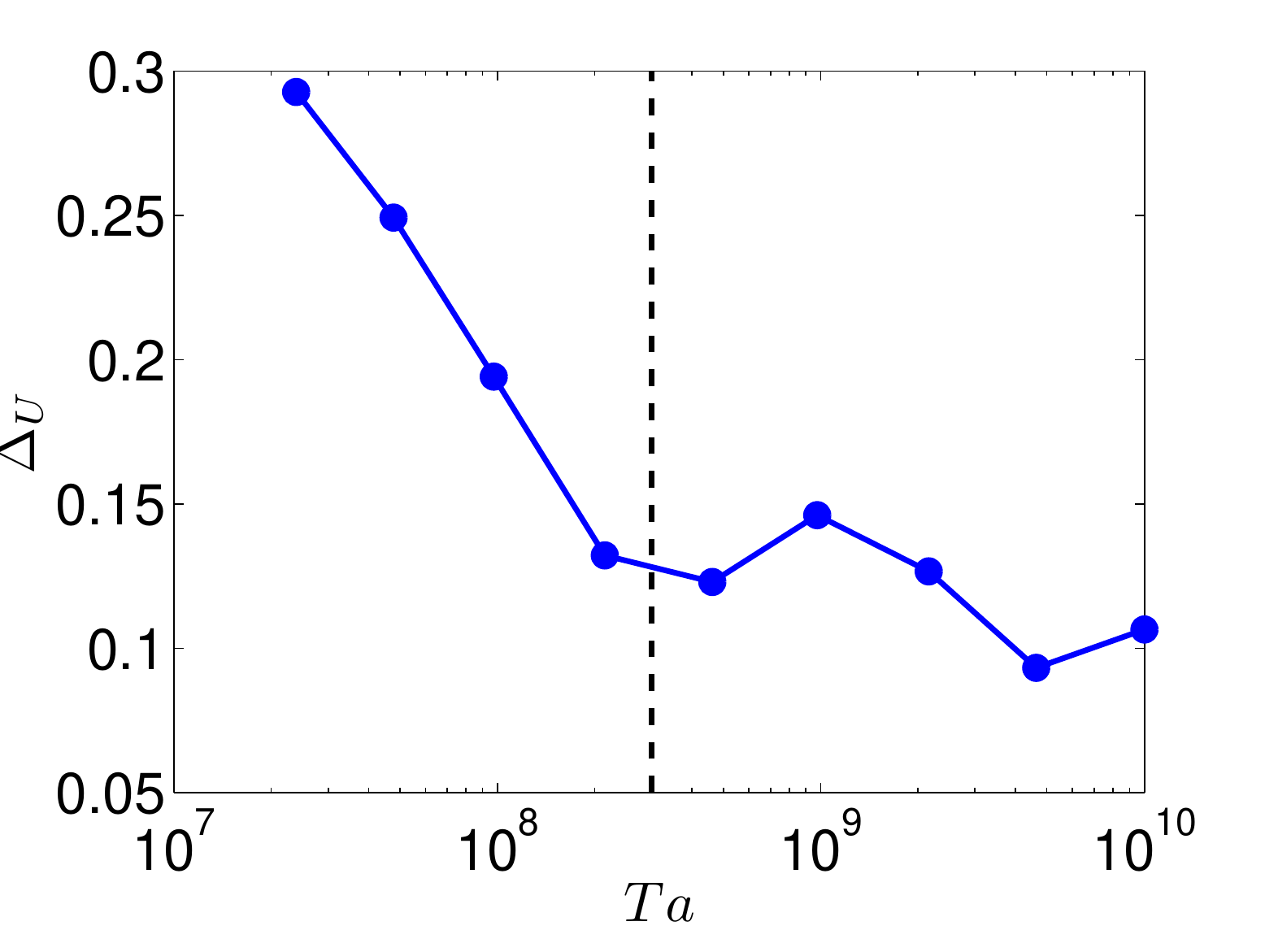}
    \caption{
Normalized velocity increment $\Delta_U=(\max_z(\bar{u}_\theta(r_a))-\min_z(u_\theta(r_a)))/\langle \bar{u}_\theta(r_a)\rangle_z$
 against $Ta$. There is a transition in the axial dependence, dropping to about $10\%$ at $Ta=3\cdot10^8$,
corresponding to the transition to the ultimate regime. 
 Remarkably, an axial dependence of about $10\%$ is still present at $Ta=10^{10}$, which corresponds to a Reynolds number of about $10^5$.
}
\label{fig:deltaufig}
\end{figure}

As a confirmation of our statement that the large scale structures are washed away, we quantify the reduction in axial dependence 
of the angular and azimuthal velocity profiles for increasing driving strength by defining the normalized velocity increment
 $\Delta_U=(\max_z(\bar{u}_\theta(r_a))-\min_z(\bar{u}_\theta(r_a)))/\langle \bar{u}_\theta(r_a)\rangle_z$. The meaning of 
$\Delta_U$ is that the larger this increment, the
stronger the axial dependence. $\Delta_U$ versus $Ta$ is shown in the right panel of figure \ref{fig:deltaufig}. As expected, the axial dependence
strongly decreases in the $Ta$-range of the transition to the ultimate regime. After the transition to the ultimate regime,
$\Delta_U$ fluctuates between $0.1$ to $0.15$, though the strength of the 
large scale wind continuously diminishes with increasing $Ta$.
Indeed, some degree of axial dependence remains, even at the highest
drivings. This result is remarkable, as even at $Ta=10^{10}$, corresponding to Reynolds numbers of $10^5$, an effect of the initial roll
state seems to remain.

\begin{figure}[ht]
    \centering
    \includegraphics[trim=0cm 0cm 0cm 0cm,width=0.47\textwidth]{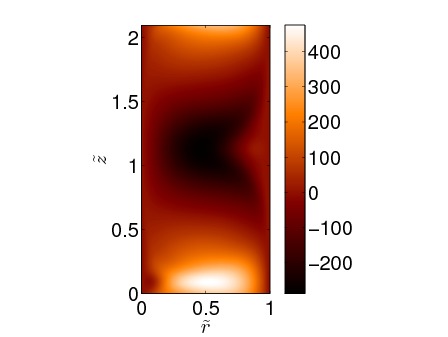}
    \includegraphics[trim=0cm 0cm 0cm 0cm,width=0.47\textwidth]{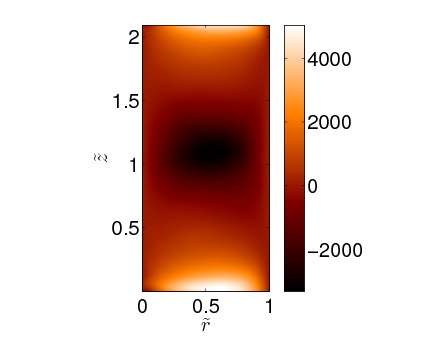}
    \caption{Contour plots of the time- and azimuthally averaged local Nusselt number $\nom(r,z)$ for (a) $Ta=4.77\cdot 10^7$ 
($\nom=8.68$)  and (b) $Ta=1.00\cdot 10^{10}$ ($\nom=51.5$). Fluctuations two orders of magnitude
 higher than the axially averaged values of $\nom$ can be
seen even at the largest drivings, indicating that remnants of the initial roll state are still present.  }
\label{fig:nuaxial}
\end{figure}

The residual axial dependence of the profiles can be understood by looking at the local 
Nusselt number $\nom(r,z)=r^3(\langle u_r\omega\rangle_{\theta,t} - \nu \partial_r \langle \omega \rangle_{\theta,t})/T_{pa}$.
Figure \ref{fig:nuaxial} shows azimuthally cut contour plots of $\nom(r,z)$ for two values of $Ta$, in the 
transitional ($Ta=4.77\cdot 10^7$, left panel) and in the ultimate regimes ($Ta=1.00\cdot 10^{10}$, right panel).
At the highest drivings, a very strong axial dependence can still be seen in $\nom$. In the bulk, $\nom$ 
is (apart from the non-dimensionalization), the correlation between $u_r$ and $\omega$.  The axial dependence in $\nom$ 
is two orders of magnitude larger than for the average values, and negative $\nom$ of even $3000$ can be seen for the largest driving. This means that
even if the azimuthal velocity loses most of its axial dependence, structures can be seen in $u_r$ (and in consequence in $u_z$)
up to Reynolds numbers of about $10^5$, which in turn is causing the residual axial dependence of the profiles $u_\theta(r)$ and $\omega(r)$.

\begin{figure}[ht]
    \centering
    \includegraphics[trim=0cm 0cm 0cm 0cm,width=0.47\textwidth]{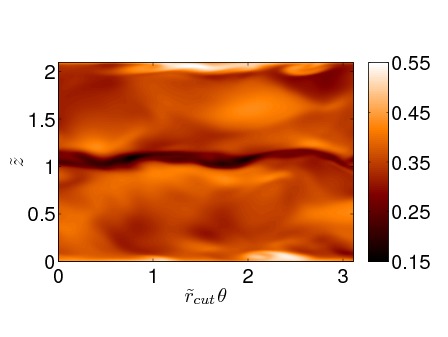}
    \includegraphics[trim=0cm 0cm 0cm 0cm,width=0.47\textwidth]{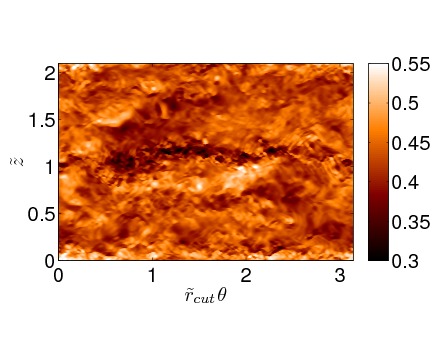}\\
    \includegraphics[trim=0cm 0cm 0cm 0cm,width=0.47\textwidth]{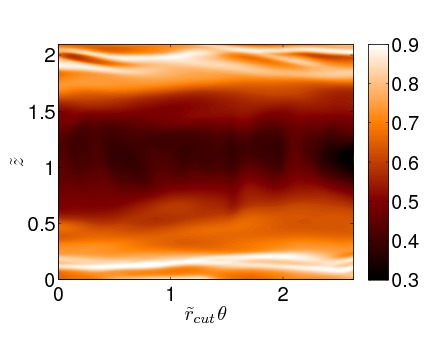}
    \includegraphics[trim=0cm 0cm 0cm 0cm,width=0.47\textwidth]{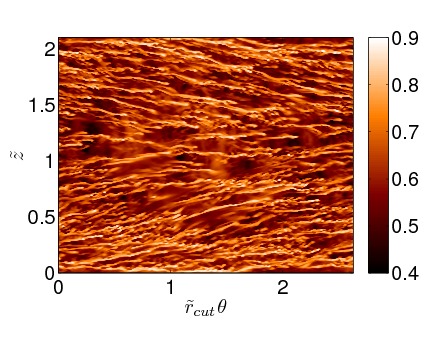}\\
    \caption{Contour plots of the instantaneous azimuthal velocity $u_\theta$ for two Taylor numbers 
at constant radius cuts (at $\tilde{r}=\tilde{r}_{cut}$) either in BL or bulk. Top-left panel: $Ta=4.77\cdot 10^7$, $\tilde{r}_{cut}=0.5$ (transition regime, bulk), 
top-right panel: $Ta=4.63\cdot 10^9$, $\tilde{r}_{cut}=0.5$ (ultimate regime, bulk), 
bottom-left panel: $Ta=4.77\cdot 10^7$, $\tilde{r}_{cut}=3.2\cdot10^{-2}$ (transition regime, BL) and
bottom-right panel: $Ta=4.63\cdot 10^9$, $\tilde{r}_{cut}=1.1\cdot10^{-2}$ (ultimate regime, BL). }
\label{fig:azimuthalutheta}
\end{figure}

Up to now, we have focused on the loss of axial dependence, and have ommited from our analysis the azimuthal structure of the flow.
This is justified by the fact that the flow is statistically homogeneous in the $\theta$-direction, so it is does not play an important role
in the transitions.
As a confirmation of this, figure \ref{fig:azimuthalutheta} shows contour plots taken at a constant radius $\tilde{r}=\tilde{r}_{cut}$ of the instantaneous 
azimuthal velocity field $u_\theta$ for two 
values of $Ta=4.77\cdot 10^7$ (transitional regime) and $Ta=4.63\cdot 10^9$ (ultimate regime) both 
in the BLs ($\tilde{r}_{cut}=3.2\cdot10^{-2}$ for $Ta=4.77\cdot 10^7$ 
and $\tilde{r}_{cut}=1.1\cdot10^{-2}$ for $Ta=4.63\cdot 10^9$) and in the bulk ($\tilde{r}_{cut}=0.5$ for both $Ta$). 

The two bottom panels in the BLs show the formation of $\omega$-plumes. These were previously interpreted to be herring-bone streaks by Dong \cite{don07}.
The axial structure of the flow present in the transitional regime can be appreciated for the two left contour plots, and it can be
seen to dissapear when looking at the right most contour plots. However, the flow shows no clear azimuthal structure in any of the panels,
and there is no indication of a flow transition if one looks at the azimuthal structure of the flow.


In conclusion, the logarithmic azimuthal velocity profile of the ultimate 
regime in TC is triggered by plume ejection that in turn is regulated by the
relative strength of the axial and radial mean flow with respect to the azimuthal one. 
This is not a finite-size effect (i.e., triggered by the upper and lower plates), as 
simulations in this study are done in a periodic domain. The transition to the ultimate 
regime in TC flow is determined by the growth of the plume ejection regions
until they cover the majority of the cylinder surface.
This occurs when the coherent wind is sufficiently weak so that the axial pressure gradient at the wall becomes
negligible, when compared to the shear from the inner cylinder, resulting in plume emission from the complete axial extent of the boundary layer.

In the ultimate regime two logarithmic boundary layers (with curvature corrections, see ref. [20])
for the angular velocities form, one at each cylinder, whose (inverse) slopes are Ta-dependent.
Also, the azimuthal velocity profiles can still reasonably well be fitted by a log-law, as done in 
experiments \cite{hui13}, though strictly speaking not both $u^+$ and $\omega^+$ can follow log-laws because
they differ by an additive constant. If $u^+$ is fitted with a log-law nonetheless, the (inverse) slopes $\kappa$ are also
Ta-dependent and thus differs from the classical von Karman constant.
In the limit of very large $Ta$, the (inverse) slopes $\kappa$ seem to tend to the universal von Karman constant $\kappa=0.41$. 
Surprisingly, in this regime some dependence on the initial
roll state can still be observed up to the highest driving achieved in these simulations.


Acknowledgments: We would like to thank H. Brauckmann, 
S. G. Huisman, C. Sun, and R.C.A. van der Veen for various stimulating discussions and for providing
the datapoints for figure \ref{fig:comparsion}.
This study is supported by ERC, FOM, and the National Computing Facilities (NCF), both sponsored by NWO. 
We acknowledge that the results of this research have been achieved using the PRACE-2IP 
project (FP7 RI-283493) resource VIP based in Germany at LRZ.

\bibliography{/home/ostillamonicor/BackedupHome/literatur}

\newpage

\appendix

\textbf{APPENDIX: NUMERICAL DETAILS}

\begin{table}[h]
  \begin{center} 
  \def~{\hphantom{0}}
  \begin{tabular}{|c|c|c|c|c|c|}
  \hline
  $Re_i$   & $Ta$   & $\nom$       & $N_\theta$x$N_r$x$N_z$ & $\eta_K/d$ & $Re_{\tau,i}$ \\
  \hline
  1.19E+04 & 2.15E+08 & 13.105 & 256x640x256  & 4.46E-03 & 4.58E+02 \\          
  1.74E+04 & 4.62E+08 & 16.940 & 256x640x512  & 3.46E-03 & 6.33E+02 \\          
  2.53E+04 & 9.75E+08 & 22.081 & 256x640x512  & 2.68E-03 & 9.37E+02 \\          
  3.76E+04 & 2.15E+09 & 29.860 & 256x640x512  & 2.04E-03 & 1.34E+03 \\          
  5.52E+04 & 4.63E+09 & 38.544 & 384x640x768  & 1.58E-03 & 1.84E+03 \\          
  8.10E+04 & 1.00E+10 & 51.554 & 512x800x1024 & 1.21E-03 & 2.60E+03 \\          
 \hline
 \end{tabular}
 \caption{This table presents the numerical results which are new to this manuscript.
For the other data points see Ostilla et al.\cite{ost13}. The first two columns show the driving, expressed as either $Re_i$ or $Ta$. The third
column shows the non-dimensionalized torque, $\nom$.
The fourth column shows the amount of grid points used in azimuthal ($N_\theta$), radial ($N_r$) and axial direction ($N_z$). The last
two columns show additional details on the grid resolution and the Kolmogorov scale $\eta_K$ in the bulk 
as well as the frictional Reynolds number $Re_\tau$ in the
boundary layers. For TC we obtain $\eta_K$ from the exact dissipation relationships as $\eta_K/d = (\sigma^{-2} Ta (\nom-1))^{-1/4}$ 
(see \cite{eck07} for a full derivation), and the frictional Reynolds number at the inner cylinder as 
$Re_{\tau,i}=u_{\tau,i} d / \nu$. All these results are for the ``reduced'' geometry
with $\Gamma=2\pi/3$ and $n_{sym}=6$.}
 \label{tbl:final}
\end{center}
\end{table}

\end{document}